\documentstyle[12pt]{article}

\begin{document}




\font\sc=cmr5 scaled\magstep1

\def\IJMP{Int.~J.~Mod.~Phys. }
\def\MPL{Mod.~Phys.~Lett. }
\def\NP{Nucl.~Phys. }
\def\PL{Phys.~Lett. }
\def\PR{Phys.~Rev. }
\def\PRL{Phys.~Rev.~Lett. }
\def\PTP{Prog.~Theor.~Phys. }
\def\ZP{Z.~Phys. }

\def\vev#1{{\langle#1\rangle}}
\def\calL{{\cal L}}
\def\aplus{{A_{+}}}
\def\aminus{{A_{-}}}
\def\Dplus{{D^{(+)}}}
\def\Dplusin{{D^{(+)}_{<}}}
\def\wight#1{\langle#1\rangle}
\def\trun#1{\langle#1\rangle_{\hbox{\sc T}}}
\def\Da{{D}}
\def\Dplusa{{D^{(+)}}}
\def\Dplusina{{D^{(+)}_{<}}}
\def\sym{{\cal S}}

\newcommand{\cev}[1]{{\stackrel{\leftarrow}{#1}}}

\def\delnl{\delta_{\hbox{\sc NL}}}
\def\ep{\epsilon}

\def\delzero{\delta_0}
\def\brs{\delta}

\def\ch{{c}}
\def\ca{{t}}
\def\cb{{c_3}}
\def\cha{{c^+}}
\def\caa{{t^+}}
\def\cba{{c_3^*}}
\def\vv{{v}}
\def\delg{\delta_{\hbox{\sc G}}}

\def\delh{{\delta_1}}
\def\dela{{\delta_2}}
\def\delb{{\delta_3}}
\def\eph{{\epsilon_1}}
\def\epa{{\epsilon_2}}
\def\epb{{\epsilon_3}}

\newcommand\gh{{\rm gh}}
\newcommand{\lb}[2]{[\![#1\,,#2]\!]}
\newcommand{\rd}{\overleftarrow{\partial}} 
\newcommand{\ld}{\overrightarrow{\partial}} 
\newcommand{\sbv}[2]{{\left(\!\left({\,{#1}\,,\,{#2}\,}\right)\!\right)}}
\newcommand{\bracket}[2]{\langle #1\,,#2\rangle}

\def\bold#1{#1\llap{$#1$\hskip.3pt}\llap{$#1$\hskip.4pt}\llap{$#1$\hskip.5pt}}
\def\bbold#1{\bold{\bold#1}}
\def\sbold#1{#1\llap{$#1$\hskip.1pt}\llap{$#1$\hskip.2pt}\llap{$#1$\hskip.3pt}}
\def\sbbold#1{\sbold{\sbold#1}}
\def\bos#1{\bbold#1}

\def\ba{\mbox{\boldmath $A$}}
\def\bb{\mbox{\boldmath $B$}}
\def\bPhi{\mbox{\boldmath $\Phi$}}
\def\bphi{\mbox{\boldmath $\phi$}}
\def\bPsi{\mbox{\boldmath $\Psi$}}
\def\bDelta{\mbox{\boldmath $\Delta$}}
\def\bdelta{\mbox{\boldmath $\delta$}}
\newcommand{\dbv}{{\Delta_{\hbox{\sc BV}}}}
\newcommand{\OmBV}{{\Omega_{\hbox{\sc BV}}}}
\newcommand{\bOmBV}{\mbox{\boldmath ${\Omega_{\hbox{\sc BV}}}$}}

\def\tx{{\cal X}}
\def\qstr{{\hat Q}}

\newcommand{\calE}{{\cal E}}
\newcommand{\calM}{{\cal M}}
\newcommand{\calX}{{\cal X}}
\newcommand{\calV}{{\cal V}}


\baselineskip 0.7cm

\begin{titlepage}
\begin{flushright}
\end{flushright}

\vskip 1.35cm
\begin{center}
{\Large \bf
Topological Field Theories and Geometry of Batalin-Vilkovisky Algebras
}
\vskip 1.2cm
Noriaki IKEDA
\footnote{ E-mail address:\ ikeda@yukawa.kyoto-u.ac.jp}
\vskip 0.4cm
{\it Ritsumeikan University \\ 
Kusatsu, Shiga 525-8577, Japan }\\
and \\
{\it Setsunan University \\ 
Neyagawa, Osaka 572-8508, Japan }
\date{}

\vskip 1.5cm

\begin{abstract}
The algebraic and geometric structures of deformations are analyzed
concerning topological field theories of Schwarz type by means of the
Batalin-Vilkovisky formalism. Deformations of the Chern-Simons-BF
theory in three dimensions induces the Courant algebroid structure on
the target space as a sigma model. Deformations of BF theories in $n$
dimensions are also analyzed. Two dimensional deformed BF theory
induces the Poisson structure and three dimensional deformed BF theory
induces the Courant algebroid structure on the target space as a sigma
model. The deformations of BF theories in $n$ dimensions induce the
structures of Batalin-Vilkovisky algebras on the target space.
\end{abstract}
\end{center}
\end{titlepage}

\setcounter{page}{2}


\rm
\section{Introduction}
\noindent
There are two types in topological field theories. 
Witten type and Schwarz type \cite{BBRT}.
The topological field theories of Schwarz type are those which
do not depend on the metric.
Roughly speaking, they are the Chern-Simons gauge theory and the BF
theories.

The author has made deformation of the BF theories in $n$ dimensions 
and Chern-Simons gauge theory coupled with the BF theory in three
dimensions in the previous papers
\cite{Ikeda:2000yq}\cite{Ikeda:2001fq}\cite{Ikeda:2002wh},
and found new topological field theories with
exotic extended gauge symmetries.
The way of analyzing those theories is
the deformation theory of the gauge theory proposed by Barnich and
Henneaux \cite{BH} \cite{BBH}.
We investigate moduli of deformations by
analyzing the BRST cohomology in the framework of Batalin-Vilkovisky (BV)
formalism (the antifield BRST formalism) \cite{Batalin:jr}. 

We consider an action $S_0[\Phi]$ of the fields $\Phi$ and its
gauge symmetry $\delzero$.
The action $S_0$ is gauge invariant under $\delzero$, that is, 
$\delzero S_0 = 0$.
The action and the gauge symmetry are deformed perturbatively as
\begin{eqnarray}
S = S_0 + g S_1 + g^2 S_2 + \cdots,
\nonumber \\
\brs \Phi = \delzero \Phi + g \delta_1 \Phi + g^2 \delta_2
\Phi + \cdots,  
\end{eqnarray}
where $g$ is a deformation parameter.
The consistency of the deformed theory requires that the
deformed action is gauge invariant under the deformed gauge symmetry,
\begin{eqnarray}
\brs S = 0,
\end{eqnarray}
and the deformed gauge symmetry is closed, that is,
\begin{eqnarray}
[\brs_{\epsilon}, \brs_{\epsilon^\prime}] 
= \brs_{[{\epsilon}, {\epsilon^\prime}]}
\label{gaugeclose}
\end{eqnarray}
holds on shell where $\epsilon$ and $\epsilon^\prime$ are gauge
parameters.
These conditions are realized as the classical
master equation, $(S, S) =0$, in the BV formalism,
where $(\cdot,\cdot)$ is the antibracket and $S$ is a BV action.
%
By deformation of the topological field theory, we mean analysis of
the freedoms of deformation of the mathematical structures defined by
the topological field theory.

The Chern-Simons gauge theory in three dimensions has the following
action:
\begin{eqnarray}
S_{{\sc CS}} =  \int_{X} \left( \frac{k_{ab}}{2} A^a \wedge d A^b
+ \frac{1}{6} f_{abc} A^a \wedge A^b \wedge A^c \right),
\label{nacs}
\end{eqnarray}
where $X$ is a three-dimensional manifold, 
$A^a$ is a one-form gauge field, 
$k_{ab} = k_{ba}$ is a nondegenerate constant and
$f_{abc}$ is the structure constant of a Lie algebra.
This theory has the following gauge symmetry:
\begin{eqnarray}
&& \brs A^a = d \epsilon^a +  k^{ab} f_{bcd} A^c \epsilon^d,
\end{eqnarray}
and the field strength $F^a$ vanishes as is seen from the equation of
motion.

The BF theory in $n$ dimensions has the following action:
\begin{eqnarray}
S = (-1)^{n-p} \int_{\Sigma}
B_{n-p-1\ a} F_{p+1}{}^a,
\end{eqnarray}
where $F_{p+1}{}^a$ is the field strength for
a $p$-form gauge field $A_p$, and 
$B_{n-p-1\ a}$ is an $n-p-1$-form auxiliary field.
The equation of motion is also, $F_{p+1}{}^a=0$.

The deformation of the Chern-Simons gauge theory 
from the abelian one has been analyzed by Barnich and
Henneaux \cite{BH}, 
who have obtained only the known nonabelian Chern-Simons gauge theory.
Deformation of the BF theory in two dimension has been analyzed by
Izawa \cite{Iz}.
There are also some similar works
\cite{Bizdadea:ic}\cite{Schwarzweller:2001pw}.
We obtain two dimensional nonlinear gauge theory (the Poisson sigma
model) \cite{II1}\cite{SS} as the deformation of the BF theory in two
dimension. 
The theory has an extended gauge symmetry.
The author has analyzed deformations of the BF theory in three
dimensions \cite{Ikeda:2000yq} and extended to those in $n$ dimensions
\cite{Ikeda:2001fq}. 
The theories obtained have some exotic gauge
symmetries which are extensions of the usual gauge symmetries of the
Lie algebras.
The deformation topological field theory of Schwarz type has been also
discussed by Edgren and Sandstrom \cite{Edgren:2002xg},
who have obtained explicit solutions in four and six dimensions.
The author has obtained a deformed topological field theory with
extended gauge symmetry for the Chern-Simons gauge theory coupled with
BF theory in three dimensions \cite{Ikeda:2002wh}.

The global version of extended gauge symmetries are trivial on the
physical S-matrix since we consider topological theories.
But the gauge symmetries of the theories are not necessarily mutually
equivalent on a topologically 
nontrivial manifold, including a manifold with boundaries.
A known example is the deformation of the BF theory in two dimensions, 
which is the nonlinear gauge theory \cite{Iz}.
We consider this as a sigma model on an 
$2$-dimensional base space $X$, mapping to a target space $M$.
This theory defines the Poisson structure on the target space $M$
as a sigma model \cite{SS}, and therefore is called 
the Poisson sigma model.
If we quantize this theory on a disc, 
correlation functions of observables on the boundary of the disc
coincide with the 
deformation quantization on the Poisson manifold $M$ \cite{CF}.
We can derive the associativity condition of 
the deformation quantization as the Ward-Takahashi identity
of the gauge symmetry.
This model has been analyzed also in the context of 
$L_\infty$-algebra \cite{LS}\cite{Sta}\cite{FLS} and of the Lie
algebroid \cite{Levin:2000fk}\cite{Olshanetsky:2002ur}.

We mainly consider such theories as a sigma model on an 
$n$-dimensional base space $X$, mapping to a target space $M$.
We consider the general deformed BF theories 
on an $n$-dimensional manifold $X$.
We analyze their
geometrical and algebraic structures induced on the target space $M$.
The $n$ dimensional topological sigma model defines a
topological open $(n-1)$-brane \cite{Park}.
Therefore we can apply analysis of deformed BF theories 
to the analysis of deformation of the topological open $(n-1)$-brane.

The key structure for analyzing higher-dimensional deformed
theories is the BV structure.
The antibracket and the BRST charge is 
geometrically reformulated as $P$-structure and $Q$-structure
\cite{Schwarz:1992nx}\cite{Alexandrov:1997kv}.
In this paper, we analyze what structures are induced on the target
space by a topological sigma
model with $P$-structure and $Q$-structure.


In section 2, we consider deformation of the Chern-Simons gauge theory
coupled with BF theory in three dimensions and discuss that this
theory has the Courant algebroid structure.
In section 3, we briefly review deformation of BF theories in $n$
dimensions. 
In section 4, we analyze the structures of deformed BF theories in
two dimensions and in three dimensions.
We point out that the theories have the Poisson structure in two
dimensions and the Courant algebroid structure in three dimensions.
We reformulate these structures in the BV formalism.                    
In section 5, we analyze the deformed BF theories in $n$ dimensions
and their Batalin-Vilkovisky algebras.

\section{Chern-Simons-BF Theory in three dimensions}
\subsection{Superfield Formalism for Chern-Simons-BF Theory}
\noindent
In this subsection, we briefly review the deformation of the
Chern-Simons gauge theory coupled with BF theory in three dimension
discussed in \cite{Ikeda:2002wh}. 

Let $X$ be a three-dimensional manifold and $M$ be an $N$-dimensional
manifold. 
$E$ denotes a vector bundle on $M$.
First we consider the abelian Chern-Simons-BF action,
\begin{eqnarray}
S_0 =  \int_{X} \left( \frac{k_{ab}}{2} A^a \wedge d A^b - B_i \wedge
d \phi^i
\right),
\label{acsbf}
\end{eqnarray}
where 
$\phi^i$ is a 0-form scalar field and is a (smooth) map from $X$ to
an $N$-dimensional target space $M$.
$A^a$ and $B_i$ are a $1$-form and a $2$-form gauge fields
respectively, and
$k_{ab}$ is a symmetric constant tensor.
We assume that $k_{ab}$ is nondegenerate and invertible.
Indices $a, b, c$ represent those on the fiber $E_x$ $(x \in X)$, and
$i, j, k$, represent the indices on the cotangent bundle $T^*M$.
The sign factor $-1$ in the front of the second term is introduced for 
convenience.

The author has analyzed all the BRST cohomology
by means of the method developed in \cite{BH} \cite{BBH}.
We obtain the following deformation of (\ref{acsbf}):
\begin{eqnarray}
S = \int_{X} \left( \frac{k_{ab}}{2} A^a \wedge d A^b 
- B_i \wedge d \phi^i 
+  f_{1a}{}^i (\phi) A^a B_i 
+  \frac{1}{6} f_{2abc} (\phi) A^a A^b A^c \right),
\label{csbf}
\end{eqnarray}
where $f_{1a}{}^i (\phi)$ and $f_{2abc} (\phi)$ satisfy the
following identities:
\begin{eqnarray}
&& k^{ab} f_{1a}{}^i (\phi) f_{1b}{}^j (\phi) = 0, 
\nonumber \\ 
&& \frac{\partial f_{1b}{}^i (\phi) }{\partial \phi^j} f_{1c}{}^j (\phi) 
- \frac{\partial f_{1c}{}^i (\phi) }{\partial \phi^j} f_{1b}{}^j (\phi) 
+ k^{ef} f_{1e}{}^i (\phi) f_{2fbc} (\phi) = 0, 
\nonumber \\
&& \left( f_{1d}{}^i (\phi) \frac{\partial f_{2abc} (\phi)}
{\partial \phi^i}
- f_{1c}{}^i (\phi) \frac{\partial f_{2dab} (\phi) }{\partial \phi^i}
+ f_{1b}{}^i (\phi) \frac{\partial f_{2cda} (\phi) }{\partial \phi^i}
- f_{1a}{}^i (\phi) \frac{\partial f_{2bcd} (\phi) }{\partial \phi^i} 
\right) \nonumber \\
&& 
+ k^{ef} (f_{2eab} (\phi) f_{2cdf} (\phi) 
+ f_{2eac} (\phi) f_{2dbf} (\phi) 
+ f_{2ead} (\phi) f_{2bcf} (\phi) )
= 0.
\label{identity3}
\end{eqnarray}
Two additional terms arise in (\ref{csbf}) 
as the most general BRST cohomology class of deformation,
where we assume that the ghost number in the action is zero
\cite{Ikeda:2002wh}.

If $f_{1a}{}^i (\phi) = 0$ and if
$f_{2abc} (\phi)$ is independent of $\phi$,
(\ref{identity3}) reduces to the usual Jacobi
identity for the Lie algebra structure constant and we have the
usual nonabelian gauge symmetry.
In general, however $f_{2abc} (\phi)$ can depend on the fields,
and thus the theory is more general than the usual 
nonabelian gauge symmetry. 

The action (\ref{csbf}) is invariant under the following 
gauge symmetry:
\begin{eqnarray}
&& \brs A^a = d c^a +  k^{ab} f_{1b}{}^i t_i
+  k^{ab} f_{2bcd} A^c c^d, \nonumber \\
&& \brs B_i = d t_i
+  \frac{\partial f_{1b}{}^j}{\partial \phi^i} (A^b t_j - c^b B_j)
+  \frac{1}{2} \frac{\partial f_{2bcd}}{\partial \phi^i} A^b A^c c^d, 
\nonumber \\
&& \brs \phi^i = - f_{1b}{}^i c^b,
\end{eqnarray}
where $c^a$ and $t_i$ are gauge parameters.
$c^a$ is a $0$-form and $t_i$ is a $1$-form.
Since this gauge algebra is an open algebra, it is appropriate to
analyze the theory in the framework of the Batalin-Vilkovisky
formalism (the antifield formalism). 

To do this, 
first we take both $\ch^a$ and $\ca_i$ to be the Grassmann odd
FP ghosts with ghost number $1$, and we introduce $\vv_i$ to be a the
Grassmann even ghost with ghost number $2$.
Next we introduce an antifield $\Phi^+$ corresponding to each field
$\Phi$.
We requires ${\rm deg}(\Phi) + {\rm deg}(\Phi^+) = n$ and
${\rm gh}(\Phi) + {\rm gh}(\Phi^+) = -1$, 
where $n$ is the dimension of the base space $X$
($n=3$ in this model),
${\rm deg}(\Phi)$ and ${\rm deg}(\Phi^+)$ are the form degrees 
of the fields $\Phi$ and $\Phi^+$, respectively
and ${\rm gh}(\Phi)$ and ${\rm gh}(\Phi^+)$ are the ghost numbers of
them. 
For functions, $F(\Phi, \Phi^+)$ and $G(\Phi, \Phi^+)$, of the fields
and the antifields,
we define the antibracket by
\begin{eqnarray}
(F, G) \equiv \frac{F \rd}{\partial \Phi} \frac{\ld G}{\partial \Phi^+}
- 
\frac{F \rd}{\partial \Phi^+} \frac{\ld G}{\partial \Phi},
\label{bracket}
\end{eqnarray}
where ${\rd}/{\partial \varphi}$ and ${\ld}/{\partial
\varphi}$ are the right and left differentiations
with respect to $\varphi$, respectively.
For two functionals $S$ and $T$, their antibracket is defined by
\begin{eqnarray}
(S, T) \equiv \int_{X}
\left(
\frac{S \rd}{\partial \Phi} \frac{\ld T}{\partial \Phi^+}
- 
\frac{S \rd}{\partial \Phi^+} \frac{\ld T}{\partial \Phi}
\right).
\label{antibracket}
\end{eqnarray}
Properties of the antibracket are summarized in the Appendix A.

In order to simplify 
notations and calculations, 
we employ the superfield formalism.
%
A superfield consists of a field, its antifield and their gauge
descendant fields.
%
For $\phi^i$, $A^{a}$ and $B_i$, 
the corresponding superfields are as follows:
\begin{eqnarray}
\bphi^i &= &
\phi^i + B^{+i} + t^{+i} + v^{+i},
\nonumber \\
\ba^a &=& 
c^{a} + A{}^{a} + k^{ab} A^+_b + k^{ab} c^+_b, 
\nonumber \\
\bb_{i} &= &
v_i + t_i + B_i + \phi^+_i.
\label{cscomponent}
\end{eqnarray}
The total degree, defined by $|F| \equiv \gh F + \deg F$, of 
the component fields belonging to each is common.
The total degrees of $\bphi^i$, 
$\ba^a$ and $\bb_i$ are $0$, $1$ and $2$, respectively.
We introduce notation the {\it dot product} (denoted by $\cdot$) for 
superfields in order to make sign factors implicit
\cite{Cattaneo:2000mc}.
The definitions and properties of the {\it dot product} are listed in
the Appendix B. 
Using the formula (\ref{dotantibra}) presented in the Appendix B,
we can rewrite (\ref{bracket}) as the antibracket on two superfields
$F$ and $G$ as follows:
\begin{eqnarray}
\sbv{F}{G} \equiv 
F \cdot \frac{\rd}{\partial \ba^a} \cdot k^{ab} 
\frac{\ld }{\partial \ba^b} \cdot G
+ 
F \cdot \frac{\rd}{\partial \bphi^i} \cdot
\frac{\ld }{\partial \bb_i} \cdot G
- 
F \cdot \frac{\rd }{\partial \bb_{i}} \cdot
\frac{\ld }{\partial \bphi^i} \cdot G.
\label{bvbracket}
\end{eqnarray}
$\sbv{\cdot}{\cdot}$ is graded symmetric and satisfy the graded Leibniz
rule and the graded Jacobi identity with respect to the total degree
of superfields. 
That is,  (\ref{bvbracket}) defines the graded Poisson bracket on superfields.
The formulae are listed in the Appendix B.
%

The action (\ref{csbf}) is extended to the BV action in terms of 
the superfields as follows:
\begin{eqnarray}
S_{BV} = \int_{X} \left( \frac{k_{ab}}{2} \ba^a \cdot d \ba^b
- \bb_i \cdot d \bphi^i
+  f_{1a}{}^i (\bphi) \cdot \ba^a \cdot \bb_i
+  \frac{1}{6} f_{2abc} (\bphi) \cdot \ba^a \cdot \ba^b \cdot \ba^c
\right),
\label{gzaction}
\end{eqnarray}
where the integration over $X$ is understood as that over the 
$3$-form part of the integrand. 
From now on, we denote this BV action $S_{BV}$ as $S$.
The gauge invariance of the action is equivalent to the following
classical master equation:
\begin{eqnarray}
\sbv{S}{S} = 0.
\label{master}
\end{eqnarray}
Substituting (\ref{gzaction}) in the condition (\ref{master}),
we obtain the identities on the structure functions
$f_{1a}{}^b(\bphi)$ and $f_{2abc}(\bphi)$ as
\begin{eqnarray}
&& k^{ab} f_{1a}{}^i \cdot f_{1b}{}^j = 0, \nonumber \\
&& \left( \frac{\ld}{\partial \bphi^j} \cdot f_{1b}{}^i \right) 
\cdot f_{1c}{}^j
- \left( \frac{\ld }{\partial \bphi^j} \cdot f_{1c}{}^i \right) 
\cdot f_{1b}{}^j
+ k^{ef} f_{1e}{}^i \cdot f_{2fbc} = 0, 
\nonumber \\
&& 
\Biggl\{ f_{1d}{}^j \cdot \left( 
\frac{\ld}{\partial \bphi^j} \cdot f_{2abc} \right)
- f_{1c}{}^j \cdot \left( \frac{\ld }{\partial \bphi^j} \cdot f_{2dab} 
\right)
+ f_{1b}{}^j \cdot \left( \frac{\ld}{\partial \bphi^j} \cdot f_{2cda}
\right)
\nonumber \\
&&
- f_{1a}{}^j \cdot \left( \frac{\ld}{\partial \bphi^j} \cdot f_{2bcd} \right)
\Biggr\} 
+ k^{ef} (f_{2eab} \cdot f_{2cdf}
+ f_{2eac} \cdot f_{2dbf}
+ f_{2ead} \cdot f_{2bcf})
= 0.
\label{stride}
\end{eqnarray}
If we set all the antifield zero in (\ref{stride}),
the identity (\ref{identity3}) is reproduced.

The BRST transformation of each field is calculated from
the definition of the BRST transformation
$\brs F = \sbv{S}{F}$ as
\begin{eqnarray}
&& \brs \ba^a = d \ba^a +  k^{ab} f_{1b}{}^j \cdot \bb_j
+  \frac{1}{2} k^{ab} f_{2bcd} \cdot \ba^c \cdot \ba^d, \nonumber \\
&& \brs \bb_i =d \bb_i
+  \left( \frac{\ld}{\partial \bphi^i} \cdot f_{1b}{}^j \right)  
\cdot \ba^b \cdot
\bb_j
+  \left( \frac{1}{6} \frac{\ld }{\partial \bphi^i} \cdot 
f_{2bcd} \right)
\cdot \ba^b \cdot \ba^c \cdot \ba^d, \nonumber \\
&& \brs \bphi^i = d \bphi^i -  f_{1b}{}^i \cdot \ba^b.
\label{brst}
\end{eqnarray}

\subsection{Courant Algebroid Structure of The CSBF Theory}
\noindent
We analyze the identities (\ref{stride}) on the structure
functions $f_1$ and $f_2$, which is equivalent to (\ref{identity3}).
The gauge algebra under this theory is the Courant algebroid.

A Courant algebroid is introduced by Courant 
in order to analyze the Dirac structure as a generalization of the 
Lie algebra of the vector fields
on the vector bundle \cite{Courant}\cite{LWX}.
A Courant algebroid is a vector bundle $\calE \rightarrow \calM$
and has a nondegenerate symmetric bilinear form
$\bracket{\cdot}{\cdot}$ 
on the bundle, a bilinear operation $\circ$ on $\Gamma(\calE)$ (the
space of sections on $\calE$), an a bundle map (called the anchor) 
$\rho: \calE \rightarrow T\calM$ satisfying the following properties
\cite{Roy01}:
\begin{eqnarray}
&& 1, \quad e_1 \circ (e_2 \circ e_3) = (e_1 \circ e_2) \circ e_3 
+ e_2 \circ (e_1 \circ e_3), \nonumber \\
&& 2, \quad \rho(e_1 \circ e_2) = [\rho(e_1), \rho(e_2)], \nonumber \\
&& 3, \quad e_1 \circ F e_2 = F (e_1 \circ e_2)
+ (\rho(e_1)F)e_2, \nonumber \\
&& 4, \quad e_1 \circ e_2 = \frac{1}{2} {\cal D} \bracket{e_1}{e_2},
\nonumber \\ 
&& 5, \quad \rho(e_1) \bracket{e_2}{e_3}
= \bracket{e_1 \circ e_2}{e_3} + \bracket{e_2}{e_1 \circ e_3},
  \label{courantdef}
\end{eqnarray}
where 
$e_1, e_2$ and $e_3$ are sections of $\calE$, and $F$ is a function on 
$\calM$;
${\cal D}$ is a map from functions on $\calM$ to $\Gamma(\calE)$ and is
defined by
$\bracket{{\cal D}F}{e} = \rho(e) F$.
Let $e^a$ be a local basis of sections of $\calE$.
Then (\ref{courantdef}) is written as
\begin{eqnarray}
&& 1, \quad e^a \circ (e^b \circ e^c) = (e^a \circ e^b) \circ e^c 
+ e^b \circ (e^a \circ e^c), \nonumber \\
&& 2, \quad \rho(e^a \circ e^b) = [\rho(e^a), \rho(e^b)], \nonumber \\
&& 3, \quad e^a \circ F e^b = F (e^a \circ e^b)
+ (\rho(e^a)F)e^b, \nonumber \\
&& 4, \quad e^a \circ e^b = \frac{1}{2} {\cal D} \bracket{e^a}{e^b},
\nonumber \\ 
&& 5, \quad \rho(e^a) \bracket{e^b}{e^c}
= \bracket{e^a \circ e^b}{e^c} + \bracket{e^b}{e^a \circ e^c}.
  \label{courantbase}
\end{eqnarray}

We consider the supermanifold ${\tilde X}$ whose grade zero part is 
a three-dimensional manifold $X$.
In our topological field theory, 
a base space $\calM$ is the space of a (smooth) map from ${\tilde X}$
to a target space $M$.
The fiber of $\calE$ is denoted as
$\calV[1]$,
where $\calV$ is a vector space and $[p]$ represents the grading shift
by $p$.
That is, the parity of the fiber of $\calE$ is reversed and the
grading of the fiber is $1$.
The local basis on $\calV[1]$ is $e^a = \ba^a$.
On this space, we define a (graded) symmetric bilinear form 
$\bracket{\cdot}{\cdot}$, a bilinear operation $\circ$ and an a
bundle map $\rho$ as follows:
\begin{eqnarray}
&& e^a \circ e^b \equiv \sbv{\sbv{S}{e^a}}{e^b}, \nonumber \\
&& \bracket{e^a}{e^b} \equiv \sbv{e^a}{e^b}, \nonumber \\
&& \rho(e^a) F(\bphi) \equiv \sbv{e^a}{\sbv{S}{F(\bphi)}}, \nonumber \\
&& {\cal D}(*) \equiv \sbv{S}{*}.
  \label{corresbase}
\end{eqnarray}
Then we can easily confirm that the gauge algebra satisfies the
conditions $1$ to $5$ of the Courant algebroid
by the identities (\ref{stride}).

Conversely, first we take the local basis $\ba^a$ on the fiber (with
the reversed parity) of the vector bundle $\calE$.
We define the graded odd Poisson structure (\ref{bvbracket})
on the bundle $\calE \oplus T^*[2]\calM$,
where the grading on the fiber direction of $T^*[2]\calM$ is shifted by
$2$.
We can take a Darboux coordinate on the antibracket
such that $\bracket{\ba^a}{\ba^b} = k^{ab}$.
We define the operations $\bracket{\cdot}{\cdot}$, $\circ$ and
$\rho$ on the basis by
\begin{eqnarray}
&& \ba^a \circ \ba^b = - k^{ac} k^{bd} f_{2cde} (\bphi) \cdot \ba^e, 
\nonumber \\
&&\bracket{\ba^a}{\ba^b} = k^{ab},
\nonumber \\
&& \rho(\ba^a) \bphi^i = - f_{1c}{}^i (\bphi) k^{ac}.
  \label{cscourant}
\end{eqnarray}
Then the conditions $1$ to $5$ of the Courant algebroid are equivalent
to the identities (\ref{stride}) on $f_1$ and $f_2$.
%
The action $S$ is the BRST charge for the Courant algebroid.
Since the master equation (\ref{master}) is equivalent to
(\ref{stride}), the relations $1$ to $5$ is represented 
by the master equation of the action $S$.


\section{Deformed BF Theory in $n$ dimensions}
\noindent
In this section, we briefly review the results of the paper
\cite{Ikeda:2001fq}, 
in which the deformed BF theory in general $n$ dimensions is 
constructed as the deformation of the abelian BF theory.
The action of the abelian BF theory in $n$ dimensions is defined as
follows: 
\begin{eqnarray}
S_0 = \sum_{p=0}^{[\frac{n-1}{2}]} \int_{X} 
(-1)^{n-p} B_{n-p-1\ a_p} d A_p{}^{a_p},
\label{abf}
\end{eqnarray}
where 
$A_p{}^{a_p}$ is a $p$-form gauge field and 
$B_{n-p-1\ a_p}$ is a $(n-p-1)$-form auxiliary field.
Indices $a_p, b_p, c_p$, etc. represent target space indices for the
$p$-from $A_p{}^{a_p}$.
Target spaces for different $p$-forms may be different.
$X$ is a base manifold on which the theory is defined.
The sign factors $(-1)^{n-p}$ are introduced for convenience.
%
%
%
%
This action has the following abelian gauge symmetry:
\begin{eqnarray}
&& \delzero A_{p}{}^{a_p} = d c_{p-1}^{(p)a_p}, \nonumber \\
&& \delzero B_{n-p-1 \ a_p} 
= d t_{n-p-2\ a_p}^{(n-p-1)},
\label{abrs}
\end{eqnarray}
where $c_{p-1}^{(p)a_p}$ is a $(p-1)$-form gauge parameter and
$t_{n-p-2\ a_p}^{(n-p-1)}$ is a $(n-p-2)$-form gauge parameter.
$(p)$ in $c_{p-1}^{(p)a_p}$ and $(n-p-1)$ in $t_{n-p-2\ a_p}^{(n-p-1)}$ 
represent that $c_{p-1}^{(p)a_p}$ is a gauge
parameter for $p$-form $A_{p}{}^{a_p}$
and $t_{n-p-2\ a_p}^{(n-p-1)}$ is one for $(n-p-1)$-form $B_{n-p-1 \ a_p}$,
respectively.
%
This gauge symmetry is reducible.
Since $A_{p}{}^{a_p}$ is a $p$-form and 
$B_{n-p-1 \ a_p}$ is a $n-p-1$-form,
we need the following towers of the 'ghost for ghosts' to analyze the
complete gauge degrees of freedom:
\begin{eqnarray}
&& \delzero A_{p}{}^{a_p} = d c_{p-1}^{(p)a_p}, 
\qquad 
\delzero B_{n-p-1 \ a_p} 
= d t_{n-p-2\ a_p}^{(n-p-1)}, \nonumber \\ 
&& \delzero c_{p-1}^{(p)a_p} = d c_{p-2}^{(p)a_p},
\qquad
\delzero t_{n-p-2\ a_p}^{(n-p-1)} 
= d t_{n-p-3\ a_p}^{(n-p-1)}, \nonumber \\
&& \vdots  \nonumber \\
&& \delzero c_{1}^{(p)a_p} = d c_0^{(p)a_p}, 
\qquad
\delzero t_{1\ a_p}^{(n-p-1)} =  d t_{0\ a_p}^{(n-p-1)},
\nonumber \\
&& \delzero c_0^{(p)a_p} = 0, 
\qquad
\delzero t_{0\ a_p}^{(n-p-1)} = 0,
\label{agauge}
\end{eqnarray}
where $c_{i}^{(p)a_p}$ are $i$-form gauge parameters and $t_{j\
a_p}^{(n-p-1)}$ are $j$-form gauge parameters.
$i = 0, \cdots, p-1$ and $j = 0, \cdots, n-p-2$.

We write the theory
in the BV formalism.
First we take $c_{i}^{(p)a_p}$ to be the
FP ghosts $i$-form with ghost number $p-i$, and $t_{j\ a_p}^{(n-p-1)}$ 
to be a $j$-form with the ghost number $n-p-1-j$.
As usual, if the ghost number is odd/even ,
the fields are Grassmann odd/even.

%
For $A_{p}{}^{a_p}$, we introduce the antifield $A_{n-p\ a_p}^{+(p)}$
, which is $(n-p)$-form with the ghost number $-1$.
For $B_{n-p-1 \ a_p}$, $B_{p+1}^{+(n-p-1)a_p}$, which is  
$(p+1)$-form with the ghost number $-1$.
For $c_{i}^{(p)a_p}$, $c_{n-i \ a_p}^{+(p)}$, which is  
$(n-i)$-form with the ghost number $-p-1+i$.
For $t_{j\ a_p}^{(n-p-1)}$, $t_{n-j}^{+(n-p-1)a_p}$, which is  
$(n-j)$-form with the ghost number $-n+p+j$.
%

%
%
For $A_{p}{}^{a_p}$ and $B_{n-p-1 \ a_p}$, 
we define corresponding superfields as
\begin{eqnarray}
\ba_p{}^{a_p} &=& 
c_0^{(p)a_p} + c_{1}^{(p)a_p} 
+ \cdots 
+ c_{p-1}^{(p)a_p} 
+ A_{p}{}^{a_p} 
+ B_{p+1}^{+(n-p-1)a_p} 
\nonumber \\ && 
+ t_{p+2}^{+(n-p-1)a_p} 
+ \cdots
+ t_{n}^{+(n-p-1)a_p}, 
\nonumber \\
%
\bb_{n-p-1\ a_p} &= &
t_{0\ a_p}^{(n-p-1)} 
+ t_{1 \ a_p}^{(n-p-1)}
+ \cdots 
+ t_{n-p-2 \ a_p}^{(n-p-1)} 
+ B_{n-p-1 \ a_p} 
+ A_{n-p \ a_p}^{+(p)} 
\nonumber \\ && 
+ c_{n-p+1 \ a_p}^{+(p)} 
+ \cdots
+ c_{n \ a_p}^{+(p)}.
\label{component}
\end{eqnarray}
The total degrees of 
$\ba_p{}^{a_p}$ and $\bb_{n-p-1\ a_p}$ are $p$ and $n-p-1$, respectively.
Since $\ba_p^{a_p}$ and $\bb_{n-p-1 \ a_p}$ are the field-antifield pair,
we can rewrite the antibracket on two superfields $F$ and $G$ 
from the definition of the antibracket (\ref{bracket}) and
superfields (\ref{component}) as follows:
\begin{eqnarray}
\sbv{F}{G} \equiv 
\sum_{p=0}^{[\frac{n-1}{2}]} 
F \cdot \frac{\rd}{\partial \ba_p{}^{a_p}} \cdot
\frac{\ld }{\partial \bb_{n-p-1 \ a_p}} \cdot G
- (-1)^{n p}
F \cdot \frac{\rd }{\partial \bb_{n-p-1 \ a_p}} \cdot
\frac{\ld }{\partial \ba_p{}^{a_p}} \cdot G.
\end{eqnarray}
$\sbv{\cdot}{\cdot}$ is graded symmetric and satisfy the graded Leibniz
rule and the graded Jacobi identity for the total degree of superfields.


The possible deformations of the BF theory in $n$
dimensions, which have been obtained in the paper
\cite{Ikeda:2001fq}, are as follows:
\begin{eqnarray}
S = S_0 + g S_1,
\label{lapS}
\end{eqnarray}
where
\begin{eqnarray}
S_0 = && \sum_{p=0}^{[\frac{n-1}{2}]} \int_{X}
(-1)^{n-p} \bb_{n-p-1\ a_p} \cdot d \ba_p{}^{a_p}, \nonumber \\
S_1 = && \sum_{p(1), \cdots, p(k), q(1), \cdots, q(l)} \int_{X} 
F_{p(1) \cdots p(k), q(1) \cdots q(l) \ a_{p(1)} \cdots a_{p(k)}}
{}^{b_{q(1)} \cdots b_{q(l)}}(\ba_0{}^{a_0})
\nonumber \\
&& \cdot \ba_{p_1}{}^{a_{p(1)}} \cdots \ba_{p_k}{}^{a_{p(k)}}
\cdot \bb_{q_1 b_{q(1)}} \cdots \bb_{q_l b_{q(l)}},
\label{s1}
\end{eqnarray}
where $F_{p(1) \cdots p(k), q(1) \cdots q(l) \ a_{p(1)} \cdots a_{p(k)}}
{}^{b_{q(1)} \cdots b_{q(l)}}(\ba_0{}^{a_0})$ 
is a function of $\ba_0{}^{a_0}$
and $p(r) \neq 0, q(s) \neq 0$ for $r = 1, \cdots, k, s = 1, \cdots, l$.
The integration over $X$ is understood 
to vanish unless the $n$-form part of the integrand.
We require that the total degree of ${\cal L}_1$ is $n$, 
that is, the ghost number of the action is zero, as in the physical 
situation, though that is not 
necessarily required by mathematical consistency of the deformations.

A necessary and sufficient condition for the theory to be
consistent is that 
the total action $S$ satisfy the following classical master
equation:
\begin{eqnarray}
\sbv{S}{S} = 0.
\label{masterb}
\end{eqnarray}
It is easily confirmed that $\brs_0 S_0 = \sbv{S_0}{S_0} = 0$ and
$\brs_0 S_1 = \sbv{S_0}{S_1} = 0$ 
if we take the proper boundary conditions so
that the integrals of total derivative
terms vanish.
Therefore the condition (\ref{master}) reduces to
\begin{eqnarray}
\sbv{S_1}{S_1} = 0.
\label{s1s1}
\end{eqnarray}
This condition imposes some identities on the structure functions 
\hfil\break
$F_{p(1) \cdots p(k), q(1) \cdots q(l) \ a_{p(1)} \cdots a_{p(k)}}
{}^{b_{q(1)} \cdots b_{q(l)}}(\ba_0{}^{a_0})$ in (\ref{s1}).

The master equation (\ref{masterb}) reduces to the equation, 
$\delzero S_1 + g/2 \sbv{S_1}{S_1} = 0$.
This is nothing but the Maurer-Cartan equation with respect to the
BRST differential $\delzero$.
We have obtained a solution of 'a flat equation' on the space of field 
theories.


The total BRST transformations $\brs$ for the superfields are
calculated as
\begin{eqnarray}
\brs \ba_p{}^a
&=&  (-1)^{n-p} \sbv{S}{\ba_p{}^a} \nonumber \\
&=&
d \ba_p{}^{a_p}
+ (-1)^{n-p} \frac{\ld }{\partial \bb_{n-p-1\ a_p}} \cdot S_1, \nonumber \\
\brs \bb_{n-p-1\ a_p}
&=& (-1)^{p(n-p)} 
\sbv{S}{\bb_{n-p-1\ a_p}} \nonumber \\
&=& 
d \bb_{n-p-1\ a_p}
+ (-1)^{p(n-p)} \frac{\ld}{\partial \ba_{p}{}^a_p} \cdot S_1.
\label{totalBRST}
\end{eqnarray}
%
%

\section{Algebraic Structures of Deformed BF Theories in Lower Dimensions}
\noindent
In this section, we analyze algebraic structures of 
deformed BF theory in lower dimensions.
\subsection{In Two Dimensions}
\noindent
First, we analyze the algebraic structure of two-dimensional
deformed BF theory as an example.


In two dimensions, (\ref{lapS}) becomes
\begin{eqnarray}
&& S = S_0 +g S_1,  \nonumber \\
&& S_0 = \int_{X}
\bb_{1 a} \cdot d \bphi^a, 
\qquad
S_1 = \int_{\Sigma} \frac{1}{2} f^{ab}(\bphi^a)
\cdot \bb_{1 a} \cdot \bb_{1 b},
\label{2daction}
\end{eqnarray}
where we rewrite notations as 
$\bphi^a = \ba_0{}^a$ and $\frac{1}{2}f^{ab}(\bphi^a) = F_{,11}(\ba_0{}^a)$.
From the condition (\ref{s1s1}),
we obtain the following identity on $f^{ab}$: 
\begin{eqnarray}
f^{cd} \cdot \frac{\ld}{\partial \bphi^d} \cdot f^{ab}
+ f^{ad} \cdot \frac{\ld}{\partial \bphi^d} \cdot f^{bc}
+ f^{bd} \cdot \frac{\ld}{\partial \bphi^d} \cdot f^{ca}= 0.
\label{2dJacobi}
\end{eqnarray}
If we set all the antifields zero, (\ref{2dJacobi}) is rewritten as
\begin{eqnarray}
f^{cd}(\phi) \frac{\partial f^{ab}(\phi)}{\partial \phi^d}
+ f^{ad}(\phi) \frac{\partial f^{bc}(\phi)}{\partial \phi^d}
+ f^{bd}(\phi) \frac{\partial f^{ca}(\phi)}{\partial \phi^d}= 0.
\label{2dJacobi2}
\end{eqnarray}
This theory is known as two-dimensional
nonlinear gauge theory (the Poisson sigma model) \cite{II1}\cite{SS}.
Under the identity (\ref{2dJacobi}), $- f^{ab}$ defines the Poisson
structure as
\begin{eqnarray}
\{F(\phi),G(\phi) \} \equiv - f^{ab}(\phi)
\frac{\partial F}{\partial \phi^a}\frac{\partial G}{\partial \phi^a},
\end{eqnarray}
on the target space $M$.
Conversely if we consider the Poisson structure $- f^{ab}$ on $M$, which
satisfies the identity (\ref{2dJacobi}), we can define the action 
(\ref{2daction}) consistently.

If we quantize this theory on a disc, 
correlation functions of observables on the boundary of the disc
coincide with the 
deformation quantization on the Poisson manifold $M$ \cite{CF}.
We can derive the associativity of the deformation quantization 
from the gauge symmetry of the theory.

The gauge symmetry of this theory is considered not as
a Lie algebra but as a Lie algebroid
\cite{Levin:2000fk}\cite{Olshanetsky:2002ur}.
A Lie algebroid is a generalization of bundles of Lie algebras over a
base manifold $\calM$.
A Lie algebroid over a manifold is a vector bundle 
$\calE \rightarrow \calM$ with a Lie algebra structure on the 
space of the sections $\Gamma(\calE)$ defined by the 
bracket $[e_1, e_2], \quad e_1, e_2 \in \Gamma(\calE)$
and a bundle map (the anchor)
$\rho: \calE \rightarrow T\calM$ satisfying the following properties:
\begin{eqnarray}
&& 1, \quad \rm{For \ any} \quad e_1, e_2 \in \Gamma(\calE), \quad
[\rho(e_1), \rho(e_2)] = \rho([e_1, e_2]),
\nonumber \\
&& 2, \quad \rm{For \ any} \quad e_1, e_2 \in \Gamma(\calE), 
\ F \in C^{\infty}(\calM), 
\nonumber \\ 
&& \qquad [e_1, F e_2] = F [e_1, e_2] + (\rho(e_1) F) e_2,
  \label{liealgdef}
\end{eqnarray}

Here we construct the bracket of the Lie algebroid from the antibracket
and the BRST structure of the theory.
In our case, $\calM$ is a space of smooth maps $\bphi^a$ from the
supermanifold  
$\tilde{X}$ whose degree zero part is a two-dimensional manifold $X$
to a target space $M$. 
A vector bundle $\calE$ is a cotangent bundle $T^*[1]\calM$, where the
grading of fiber 
direction is shifted by one.
The Lie bracket of two sections $e_1$ and $e_2$ is defined by
\begin{eqnarray}
[e_1, e_2] \equiv \sbv{\sbv{S}{e_1}}{e_2},
\end{eqnarray}
and the anchor is defined by
\begin{eqnarray}
\rho(e) F(\bphi) \equiv \sbv{e}{\sbv{S}{F(\bphi)}}.
\end{eqnarray}
Then $[e_1, e_2] = - [e_2, e_1]$ is confirmed from the graded Jacobi 
identity of the antibracket and $\sbv{e_1}{e_2}=0$.
A Lie algebroid conditions 1 and 2 on the bracket $[\cdot, \cdot]$ and the
anchor map $\rho$ is obtained from the properties of the
antibracket.
The above definition defines
the following ``noncommutative'' relation on the coordinates:
\begin{eqnarray}
[\bphi^a, \bphi^b] = - f^{ab}(\bphi),
\end{eqnarray}
and
the anchor is a differentiation on functions of $\bphi$ as
\begin{eqnarray}
\rho(\bphi^a) F(\bphi) = - f^{ab}(\bphi) \cdot
\frac{\ld}{\partial \bphi^b} \cdot F(\bphi).
\end{eqnarray}

\subsection{In Three Dimensions}
\noindent
We consider the deformed BF theory in three dimensions.
In this case, the theory defines the topological open 2-brane as a
sigma model \cite{Park}.
The total action (\ref{lapS}) becomes as follows:
\begin{eqnarray}
&& S = S_0 +g S_1,  \nonumber \\
&& S_0 = \int_{X}
[- \bb_{2 i} \cdot d \bphi{}^i + \bb_{1 a} \cdot d \ba_1{}^a], 
\nonumber \\
&& S_1 = \int_{X} 
[f_1{}_a{}^i(\bphi) \cdot \ba_1{}^a \cdot \bb_{2 i} 
+ f_2^{ib}(\bphi) \cdot \bb_{2 i} \cdot \bb_{1 b}
+ \frac{1}{3!} f_{3abc}(\bphi) \cdot \ba_1{}^a \cdot \ba_1{}^b 
\cdot \ba_1{}^c
\nonumber \\
&& 
+ \frac{1}{2} f_{4ab}{}^c(\bphi) \cdot \ba_1{}^a \cdot \ba_1{}^b 
\cdot \bb_{1 c}
+ \frac{1}{2} f_{5a}{}^{bc}(\bphi) \cdot \ba_1{}^a \cdot \bb_{1 b}
\cdot \bb_{1 c}
\nonumber \\
&& 
+ \frac{1}{3!} f_6{}^{abc}(\bphi) \cdot \bb_{1 a} \cdot \bb_{1 b}
\cdot \bb_{1 c}],
\label{3daction}
\end{eqnarray}
where we set
$f_1{}_a{}^i = F_{1,2 a}{}^i$,
$f_2^{ib} = F_{,21}{}^{ib}$,
$\frac{1}{3!}f_{3abc} = F_{111, abc}$,
$\frac{1}{2}f_{4ab}{}^c = F_{11,1 ab}{}^c$,
$\frac{1}{2}f_{5a}{}^{bc} = F_{1,11 a}{}^{bc}$,
$\frac{1}{3!}f_6{}^{abc} = F_{,111 }{}^{abc}$,
for clarity. 
The condition of the classical
master equation (\ref{s1s1}) imposes the following identities on six
$f_i$'s, $i=1, \cdots, 6$\cite{Ikeda:2000yq}:
\begin{eqnarray}
&& 
f_{1}{}_e{}^i \cdot f_{2}{}^{je} + f_{2}{}^{ie} \cdot f_{1}{}_e{}^j =
0,
\nonumber \\
&& 
\left( \frac{\ld}{\partial \bphi^j} \cdot f_{1}{}_c{}^i 
\right) \cdot f_{1}{}_b{}^j 
- \left( \frac{\ld}{\partial \bphi^j} \cdot f_{1}{}_b{}^i
\right) \cdot f_1{}_c {}^j
+ f_1{}_e{}^i \cdot f_{4bc}{}^e + f_{2}{}^{ie} \cdot f_{3ebc} = 0, 
\nonumber \\
&&
- f_1{}_b{}^j  \cdot \left( \frac{\ld}{\partial \bphi^j} \cdot
f_{2}{}^{ic} \right)
+ f_{2}{}^{jc} \cdot \left( \frac{\ld}{\partial \bphi^j}  
\cdot f_1{}_b{}^i \right)
+ f_1{}_e{}^i \cdot f_{5b}{}^{ec} - f_{2}{}^{ie} \cdot 
f_{4eb}{}^c = 0, 
\nonumber \\
&&
f_{2}{}^{jb} \cdot \left( \frac{\ld}{\partial \bphi^j} \cdot
f_{2}{}^{ic} \right)
- f_{2}{}^{jc} \cdot \left( \frac{\ld}{\partial \bphi^j} 
\cdot f_{2}{}^{ib} \right)
+ f_1{}_e{}^i \cdot f_{6}^{ebc} + f_{2}{}^{ie} \cdot f_{5e}{}^{bc} = 0, 
\nonumber \\
&&
f_1{}_{[a}{}^j \cdot \left( \frac{\ld}{\partial \bphi^j} 
\cdot f_{4bc]}{}^d \right)
- f_{2}{}^{jd} \cdot \left( \frac{\ld}{\partial \bphi^j} 
\cdot f_{3abc} \right)
+ f_{4e[a}{}^d \cdot f_{4bc]}{}^{e} + f_{3e[ab} \cdot f_{5c]}{}^{de} =
0,
\nonumber \\
&&
f_1{}_{[a}{}^j \cdot \left( \frac{\ld}{\partial \bphi^j} \cdot
f_{5b]}{}^{cd} \right)
+ f_{2}{}^{j[c} \cdot \left( \frac{\ld}{\partial \bphi^j} \cdot 
f_{4ab}{}^{d]} \right)
\nonumber \\
&& 
+ f_{3eab} \cdot f_6{}^{ecd} 
+ f_{4e[a}{}^{[d} \cdot f_{5b]}{}^{c]e} 
+ f_{4ab}{}^e \cdot f_{5e}{}^{cd} = 0, 
\nonumber \\
&&
f_1{}_a{}^j \cdot \left( \frac{\ld}{\partial \bphi^j} 
\cdot f_{6}{}^{bcd} \right)
- f_{2}{}^{j[b} \cdot \left( \frac{\ld}{\partial \bphi^j} 
\cdot f_{5a}{}^{cd]} \right)
+ f_{4ea}{}^{[b} \cdot f_6{}^{cd]e} + f_{5e}{}^{[bc} \cdot
f_{5a}{}^{d]e} = 0, 
\nonumber \\
&&
f_{2}{}^{j[a} \cdot \left( \frac{\ld}{\partial \bphi^j} \cdot
f_{6}{}^{bcd]} \right)
+ f_6{}^{e[ab} \cdot f_{5e}{}^{cd]} = 0, 
\nonumber \\
&&
f_1{}_{[a}{}^j \cdot \left( \frac{\ld}{\partial \bphi^j} 
\cdot f_{3bcd]} \right)
+ f_{4[ab}{}^{e} \cdot f_{3cd]e} = 0,
\label{3dJacobi}
\end{eqnarray}
where $[\cdots]$ on the indices represents the antisymmetrization for
them, e.g., $\Phi_{[ab]} = \Phi_{ab} - \Phi_{ba}$.

If we set all the antifields zero, we obtain the ordinary action
without antifields as follows:
\begin{eqnarray}
&& S = S_0 +g S_1,  \nonumber \\
&& S_0 = \int_{X} \left[ - B_{2\ i} d \phi^i 
+ B_{1\ a} d A_1{}^a \right], \nonumber \\ 
&& S_1 = \int_{X} 
[f_1{}_a{}^i(\phi) A_1{}^a B_{2\ i} 
+ f_2{}^{ib}(\phi) B_{2 i} B_{1 b}
+ \frac{1}{3!} f_{3abc}(\phi) A_1{}^a A_1{}^b A_1{}^c
\nonumber \\
&& 
+ \frac{1}{2} f_{4ab}{}^c(\phi) A_1{}^a A_1{}^b B_{1 c}
+ \frac{1}{2} f_{5a}{}^{bc}(\phi) A_1{}^a B_{1 b} B_{1 c}
+ \frac{1}{3!} f_6{}^{abc}(\phi) B_{1 a} B_{1 b} B_{1 c}].
\label{3dbf}
\end{eqnarray}
And (\ref{3dJacobi}) reduces to the following identities:
\begin{eqnarray}
&& 
f_{1}{}_e{}^i f_{2}{}^{je} + f_{2}{}^{ie} f_{1}{}_e{}^j = 0, 
\nonumber \\
&& 
\frac{\partial f_{1}{}_c{}^i}{\partial \phi^j} f_{1}{}_b{}^j
- \frac{\partial f_{1}{}_b{}^i}{\partial \phi^j} f_1{}_c {}^j
+ f_1{}_e{}^i f_{4bc}{}^e + f_{2}{}^{ie} f_{3ebc} = 0, 
\nonumber \\
&&
- f_1{}_b{}^j \frac{\partial f_{2}{}^{ic}}{\partial \phi^j} 
+ f_{2}{}^{jc} \frac{\partial f_1{}_b{}^i}{\partial \phi^j} 
+ f_1{}_e{}^i  f_{5b}{}^{ec} - f_{2}{}^{ie} f_{4eb}{}^c = 0, 
\nonumber \\
&&
f_{2}{}^{jb} \frac{\partial f_{2}{}^{ic}}{\partial \phi^j} 
- f_{2}{}^{jc} \frac{\partial f_{2}{}^{ib}}{\partial \phi^j} 
+ f_1{}_e{}^i f_{6}^{ebc} + f_{2}{}^{ie} f_{5e}{}^{bc} = 0, 
\nonumber \\
&&
f_1{}_{[a}{}^j \frac{\partial f_{4bc]}{}^d}{\partial \phi^j} 
- f_{2}{}^{jd} \frac{\partial f_{3abc}}{\partial \phi^j} 
+ f_{4e[a}{}^d f_{4bc]}{}^{e} + f_{3e[ab} f_{5c]}{}^{de} = 0, 
\nonumber \\
&&
f_1{}_{[a}{}^j \frac{\partial f_{5b]}{}^{cd}}{\partial \phi^j} 
+ f_{2}{}^{j[c} \frac{\partial f_{4ab}{}^{d]}}{\partial \phi^j} 
+ f_{3eab} f_6{}^{ecd} 
+ f_{4e[a}{}^{[d} f_{5b]}{}^{c]e} + f_{4ab}{}^e f_{5e}{}^{cd} = 0, 
\nonumber \\
&&
f_1{}_a{}^j \frac{\partial f_{6}{}^{bcd}}{\partial \phi^j} 
- f_{2}{}^{j[b} \frac{\partial f_{5a}{}^{cd]}}{\partial \phi^j} 
+ f_{4ea}{}^{[b} f_6{}^{cd]e} + f_{5e}{}^{[bc} f_{5a}{}^{d]e} = 0, 
\nonumber \\
&&
f_{2}{}^{j[a} \frac{\partial f_{6}{}^{bcd]}}{\partial \phi^j} 
+ f_6{}^{e[ab} f_{5e}{}^{cd]} = 0, 
\nonumber \\
&&
f_1{}_{[a}{}^j \frac{\partial f_{3bcd]}}{\partial \phi^j} 
+ f_{4[ab}{}^{e} f_{3cd]e} = 0.
\label{3dJacobi2}
\end{eqnarray}


We consider the supermanifold ${\tilde X}$ which grade zero part is 
a three-dimensional manifold $X$.
In our topological field theory, 
a base space $\calM$ is the space of a (smooth) map from ${\tilde X}$
to a target space $M$.
We consider the same setting in the section 2.
The parity of the fiber of $\calE$ is reversed and the grading is
shifted by one.
The fiber is $\calV[1] \oplus \calV^*[1]$,
where $\calV$ is a vector space and $[p]$ represents the grading
shifted by $p$.
We introduce an graded odd Poisson bracket (the antibracket)
$\sbv{\cdot}{\cdot}$ on the space. 
We take a local basis on $\Gamma(\calE)$ as 
$e^a = \ba_{1}{}^a, \bb_{1\ a}$, which are 
Darboux coordinates
such that $\sbv{\ba_1{}^a}{\ba_1{}^b} = 
\sbv{\bb_{1a}}{\bb_{1b}} = 0$ and
$\sbv{\ba_1{}^a}{\bb_{1b}} = \delta^{a}{}_b$.
In other words, $\ba_{1}{}^a$ and $\bb_{1\ a}$ are BV field-antifield
pairs. 

We define a graded symmetric bilinear form 
$\bracket{\cdot}{\cdot}$, a bilinear operation $\circ$ and an a
bundle map $\rho$ from the antibracket as follows:
\begin{eqnarray}
&& e^a \circ e^b \equiv \sbv{\sbv{S}{e^a}}{e^b}, \nonumber \\
&& \bracket{e^a}{e^b} \equiv \sbv{e^a}{e^b}, \nonumber \\
&& \rho(e^a) F(\bphi) \equiv \sbv{e^a}{\sbv{S}{F(\bphi)}}, \nonumber \\
&& {\cal D}(*) \equiv \sbv{S}{*}.
  \label{corresbase2}
\end{eqnarray}
Then we can confirm that the gauge algebra satisfies the
conditions $1$ to $5$ of the Courant algebroid defined in section 2
from the identity (\ref{3dJacobi}) on structure functions $f$'s.
This theory is also considered as a generalization of the model 
in the section 2. 

Conversely, first we define the graded odd Poisson structure
$\bracket{\cdot}{\cdot}$
on the bundle $\calE \oplus T^*[2]\calM$,
where the grading on the fiber direction of $T^*[2]\calM$ is shifted
by $2$.
We define the operations 
$\circ$ and $\rho$ on the basis as follows:
\begin{eqnarray}
&& \ba_1{}^a \circ \ba_1{}^b = - f_{5c}{}^{ab}(\bphi) \cdot \ba_1{}^c
- f_{6}{}^{abc}(\bphi) \cdot \bb_{1c}, \nonumber \\
&& \ba_1{}^a \circ \bb_{1b} = - f_{4bc}{}^{a}(\bphi) \cdot \ba_1{}^c
+ f_{5b}{}^{ac}(\bphi) \cdot \bb_{1c}, \nonumber \\
&& \bb_{1a} \circ \bb_{1b} = - f_{3abc}(\bphi) \cdot \ba_1{}^c
- f_{4ab}{}^{c}(\bphi) \cdot \bb_{1c}, \nonumber \\
&& \rho(\ba_1{}^a) \bphi^i = - f_{2}{}^{ia}(\bphi),
\nonumber \\
&& \rho(\bb_{1a}) \bphi^i = - f_{1a}{}^i(\bphi).
  \label{bfcourant}
\end{eqnarray}
Then the conditions $1$ to $5$ of the Courant algebroid are equivalent
to the identities (\ref{3dJacobi}) on six $f$'s.
%
Since the master equation (\ref{master}) is equivalent to
(\ref{3dJacobi}), the relations $1$ to $5$ is equivalent to 
the master equation of the action $S$.

The theory define the Courant algebroid structure on 
the target space as a sigma model.
Therefore we can consider this model as 'the Courant sigma model'.
We find that the topological open 2-brane has the Courant algebroid
structure.

\newcommand{\taubracket}[2]{\tau( #1\,,#2)}
\section{Algebraic Structure in $n$ Dimensions}
\noindent
We discuss the structures of $n$-dimensional deformed BF theories in
this section.
The Batalin-Vilkovisky (antifield BRST) formalism is the key device
to analyze general deformed BF theories.

A graded supermanifold $M$ with nonnegative integer grading is an
$N$-manifold if the
integer grading is compatible with parity. 
This means that bosonic fields have even weights and
fermionic fields have odd weights \cite{Roy01}\cite{Sev}.
Our superfields satisfy this condition.
A $P$-manifold is defined as a supermanifold equipped with an odd 
non-degenerate closed 2-form.
This 2-form defines the odd Poisson bracket, which
is nothing but the antibracket.
A $Q$-manifold is a
supermanifold endowed with an degree
$+1$ vector field $Q$ whose square is zero, $Q^2 =0$. 
$Q$ is nothing but the BRST charge and realized by the BV action.
A $QP$-manifold is defined as a $Q$-manifold with an odd symplectic
structure which is $Q$-invariant.
The antibracket and the BV action as the solution of the master
equation define a $QP$-structure
\cite{Alexandrov:1997kv}.
The sigma models based on our topological field theories 
have the $N$$P$$Q$-structures, 
and induce the geometry with the $N$$P$$Q$-structures on the target
space. 

Let $X$ be a $n$-dimensional manifold and let $\tilde X$ be a
supermanifold whose grade zero part is a $n$-dimensional manifold $X$.
A base space $\calM$ is the space of a (smooth) map $\bphi^{a_0} =
\ba_0{}^{a_0}$ from ${\tilde X}$ to a target space $M$.
The fiber of $\calE$ is graded.
The fiber is $\oplus_{p=1}^{[(n-1)/2]}(\calV_p[p] \oplus \calV^*_p[n-p-1])$,
where $\calV_p$, $p = 1, \cdots, [(n-1)/2]$ are vector spaces and
$[p]$ in $\calV_p[p]$ represents the grading shifted by $p$.
$[n-p-1]$ in $\calV^*_p[n-p-1]$ represents the grading shift by $n-p-1$.
Grading is $1$ to $n-2$ and compatible with the parity.

${\Sigma}_n$-manifold is an $NQ$-manifold with a $Q$-invariant
symplectic form of degree $n$ \cite{Sev}.
In our model, the ${\Sigma}_{n-1}$-structure is realized by  
the $n$-dimensional topological field theory.

We take a local basis on $\Gamma(\calE)$ as 
$e^a = \ba_p{}^{a_p}, \bb_{n-p-1\ a_p}$, where $p \neq 0$.
We introduce a graded odd Poisson bracket (the antibracket) 
$\sbv{\cdot}{\cdot}$ on the
space $\calE \oplus T^*[n-1]\calM$,
where the grading on the fiber direction of $T^*[n-1]\calM$ is shifted by
$n-1$.
$\ba_p{}^{a_p}$ and $\bb_{n-p-1\ a_p}$ are 
Darboux coordinates such that 
\begin{eqnarray}
&& \sbv{\ba_p{}^{a_p}}{\ba_q{}^{b_q}} =
\sbv{\bb_{n-p-1 \ a_p}}{\bb_{n-q-1 \ b_q}} = 0, 
\nonumber \\
&& \sbv{\ba_p{}^{a_p}}{\bb_{n-q-1 \ b_q}} 
= \delta^{p}{}_{q} \delta^{a_p}{}_{b_q}.
\end{eqnarray}
We define the following three operations:
\begin{eqnarray}
&& \bracket{E_1}{E_2}
\equiv
\sbv{E_1}{E_2},
\\
&& \taubracket{E_1}{E_2}
\equiv
\sbv{\sbv{S}{E_1}}{E_2},
\\
&& {\cal D}(*)
\equiv
\sbv{S}{*},
\end{eqnarray}
where $E_1, E_2 \in \Gamma(\calE)$ or $\in C^{\infty}(\calM)$.
Note that the degree of $\bracket{\cdot}{\cdot}$ is $-n+1$, 
the degree of $\taubracket{\cdot}{\cdot}$ is $-n+2$, 
and the degree of ${\cal D}$ is $1$.
${\cal D}$ is a differentiation.
We can generalize three operations on functions of $\ba_p{}^{a_p}$,
$p = 0, \cdots, [(n-1)/2]$, and
$\bb_{n-q-1\ a_q}$, $p = 1, \cdots, [(n-1)/2]$. 
$\bracket{E_1}{E_2}$ is a 
graded symmetric bilinear form from the property of the antibracket as
follows:
\begin{eqnarray}
\bracket{E_1}{E_2} = -(-1)^{(|E_1| + 1 - n)(|E_2| + 1 - n)} 
\bracket{E_2}{E_1}.
\end{eqnarray}
Three operations are not independent. In fact, the following
identity is satisfied:
\begin{eqnarray}
\bracket{{\cal D}E_1}{E_2} 
= \taubracket{E_1}{E_2},
\label{brataud}
\end{eqnarray}
because both sides are equal to $\sbv{\sbv{S}{E_1}}{E_2}$.
We can prove the following identities including 
the graded symmetric property of $\bracket{\cdot}{\cdot}$, derivation
properties and the Jacobi identities for
$\bracket{\cdot}{\cdot}$ and $\taubracket{\cdot}{\cdot}$, from the
properties of the antibracket and the BV action $S$ of the
deformed BF theories: 
\begin{eqnarray}
&& \bracket{E_1}{E_2} = -(-1)^{(|E_1| + 1 - n)(|E_2| + 1 - n)} 
\bracket{E_2}{E_1},
\\
&& \bracket{E_1}{E_2 \cdot E_3} = \bracket{E_1}{E_2} \cdot E_3
+ (-1)^{(|E_1| + 1 - n)|E_2|} E_2 \cdot \bracket{E_1}{E_3},
\\
&& (-1)^{(|E_1| + 1 - n)(|E_3| + 1 - n)} \bracket{E_1}{\bracket{E_2}{E_3}}
+ {\rm cyclic \ permutations} = 0.
\label{brajacobi} \\
&& \taubracket{E_1}{E_2 \cdot E_3}
= \taubracket{E_1}{E_2} \cdot E_3
+ (-1)^{(|E_1| - n)|E_2|}
E_2 \cdot \taubracket{E_1}{E_3},
\label{taulib} \\
&& 
\taubracket{E_1}{\bracket{E_2}{E_3}} 
= \bracket{\taubracket{E_1}{E_2}}{E_3}
+ (-1)^{(|E_1| - n)(|E_2| +1 - n)} \bracket{E_2}{\taubracket{E_1}{E_3}},
\label{taubrajacobi} \\
&& \taubracket{E_1}{\taubracket{E_2}{E_3}} 
\nonumber \\
&& \qquad
= (-1)^{(|E_1| - n)} \taubracket{\taubracket{E_1}{E_2}}{E_3}
+ (-1)^{(|E_1| - n)(|E_2| - n)} \taubracket{E_2}{\taubracket{E_1}{E_3}}.
\label{taujacobi}
\end{eqnarray}
Generally, $\taubracket{E_1}{E_2}$ is not graded antisymmetric as 
\begin{eqnarray}
\taubracket{E_1}{E_2}
= (-1)^{(|E_1| +1 - n)(|E_2| + 1 - n)} \taubracket{E_2}{E_1}
+ {\cal D} \bracket{E_1}{E_2}.
\label{taucom}
\end{eqnarray}
If the last term ${\cal D} \bracket{E_1}{E_2}$ vanish,
$\taubracket{\cdot}{\cdot}$ is graded antisymmetric.
We can define a graded antisymmetric bracket $[E_1, E_2]$ as follows: 
\begin{eqnarray}
\taubracket{E_1}{E_2} = [E_1, E_2] 
+ \frac{1}{2} {\cal D} \bracket{E_1}{E_2},
\end{eqnarray}
or
\begin{eqnarray}
[E_1, E_2] \equiv \frac{1}{2} \{\taubracket{E_1}{E_2} + 
(-1)^{(|E_1| + 1 - n)(|E_2| + 1 - n)}  \taubracket{E_1}{E_2} 
\}.
\end{eqnarray}
In fact, $[\cdot, \cdot]$ is graded antisymmetric as
\begin{eqnarray}
[E_2, E_1] = (-1)^{(|E_1| + 1 - n)(|E_2| + 1 - n)} [E_1, E_2],
\end{eqnarray}
However in the theory in dimensions higher than two,
this graded antisymmetric bracket does not generally satisfy the
graded Jacobi identity. 
It is understood as breaking of the Jacobi identity on the 
Courant algebroid in three dimensions.
The operations which satisfy the Jacobi identities are
$\taubracket{\cdot}{\cdot}$ and $\bracket{\cdot}{\cdot}$.

In the deformed BF theory in two dimensions, $E$ is a function of
$\bphi^a$. 
We can confirm that the algebra is the same one in the section 4.1.
We find $\bracket{E_1}{E_2}=0$ and $\taubracket{E_1}{E_2} = [E_1,
E_2]$.
The anchor is essentially the same with $\taubracket{\cdot}{\cdot}$, 
since $[E_1, E_2] = - \rho(E_2) E_1$.

In three-dimensional deformed BF theory, $E$ is expanded by $\bphi^i, 
\ba_{1}{}^a$ and $\bb_{1a}$.
$\bracket{E_1}{E_2}=0$ if $E_1$ or $E_2$ is a function of $\bphi^i$.
If $E_1$ and $E_2$ are sections of $\calE$, 
$\bracket{E_1}{E_2}$ is equal to $\bracket{\cdot}{\cdot}$ 
in section 4.2.
The $\tau$ operation satisfies that 
$\taubracket{F(\bphi)}{G(\bphi)} =0$,
$\taubracket{F(\bphi)}{e} = \rho(e) F(\bphi)$ and 
$\taubracket{e_1}{e_2} = e_1 \circ e_2$, 
where $e$, $e_1$ and $e_2$ are sections of $\calE$.
If we substitute the relations into the above, we can confirm the five
conditions 
(\ref{courantdef}) of the Courant algebroid from the identities
(\ref{taulib}) -- (\ref{taucom}).
The condition 1 of the Courant algebroid is the graded Jacobi
identity (\ref{taujacobi}). 
If one of $E$ is a function of $\bphi^i$ in (\ref{taujacobi}),
we obtain the condition 2. 
The condition 3 follows from (\ref{taulib}) if 
$E_2$ is a function of $\bphi^i$.
The condition 4 is obtained from (\ref{taucom}), 
the condition 5 from (\ref{taubrajacobi}).

Conversely, we assume $\bracket{E_1}{E_2}$, 
$\taubracket{\cdot}{\cdot}$ and ${\cal D}$ on $\Gamma(\calE)$
which satisfy (\ref{brataud}) to (\ref{taujacobi}) on an $N$-manifold
$\calE \oplus T^*[n-1]\calM$, where
the degree of $\bracket{\cdot}{\cdot}$ is $-n+1$, 
the degree of $\taubracket{\cdot}{\cdot}$ is $-n+2$, 
and the degree of ${\cal D}$ is $1$.
If at least one of $E$ is a function of $\bphi$, 
$\taubracket{E_1}{E_2}$ is considered as the anchor.
$\taubracket{F(\bphi)}{G(\bphi)} = 0$ if $n \geq 3$ because
we consider $N$-manifold with 'nonnegative' integer degree and 
the degree, $-n+2$, of the left hand side is negative.
Then we can prove ${\cal D}^2=0$ is equivalent with the compatibility of
(\ref{brataud}), the Jacobi identities (\ref{taujacobi}) and 
(\ref{brajacobi}).
Therefore the algebra defines the $QP$-structure on the target space.
$Q$-structure is $\bracket{E_1}{E_2}$ and $P$-structure is ${\cal D}$.

$\bracket{\cdot}{\cdot}$ is the graded odd Poisson bracket 
on $\calE \oplus T^*\calM$,
therefore we can find the Hamiltonian $S$ for the vector field ${\cal
D}$ such that ${\cal D} E = \bracket{S}{E}$.
The solution is our BV action.
${\cal D}^2=0$ is equivalent to the classical master equation 
$\bracket{S}{S}=0$.

The algebroid with
$\bracket{E_1}{E_2}$, $\taubracket{\cdot}{\cdot}$ and ${\cal D}$ 
which satisfies the identities (\ref{brataud}) to (\ref{taujacobi})
is the Batalin-Vilkovisky algebra based 
on the topological field theory in $n$ dimensions.
The deformed BF theory in $n$ dimensions defines the above
algebroid structures on the target space as 
'a Batalin-Vilkovisky sigma model'.

\section*{Acknowledgments}
The author would like to thank Thomas Strobl, Jim Stasheff and Ping Xu
for valuable discussions, comments and advice.  
He would like to thank Noboru Nakanishi for careful reading of the
manuscript.
\newcommand{\bibit}{\sl}

\section*{Appendix A, Antibracket}
In $n$ dimensions, 
we define the antibracket for functions $F(\Phi, \Phi^+)$ and $G(\Phi,
\Phi^+)$ of the fields and the antifields
as follows:
\begin{eqnarray}
(F, G) \equiv \frac{F \rd}{\partial \Phi} \frac{\ld G}{\partial \Phi^+}
- 
\frac{F \rd}{\partial \Phi^+} \frac{\ld G}{\partial \Phi},
\label{anti}
\end{eqnarray}
where ${\rd}/{\partial \varphi}$ and ${\ld}/{\partial
\varphi}$ are the right differentiation and the left differentiation
with respect to $\varphi$, respectively.
The following identity about left and right derivative is useful,
\begin{eqnarray}
\frac{\ld F}{\partial \varphi} =  (-1)^{(\gh F - \gh \varphi) \gh \varphi 
+ (\deg F - \deg \varphi) \deg \varphi}
\frac{F \rd}{\partial \varphi}.
\label{lrdif}
\end{eqnarray}
For two functionals $S$ and $T$, the antibracket is defined as follows:
\begin{eqnarray}
(S, T) \equiv \int_{X}
\left(
\frac{S \rd}{\partial \Phi} \frac{\ld T}{\partial \Phi^+}
- 
\frac{S \rd}{\partial \Phi^+} \frac{\ld T}{\partial \Phi}.
\right)
\label{antif}
\end{eqnarray}
The antibracket satisfies the following identities:
\begin{eqnarray}
&& (F, G) = -(-1)^{(\deg F - n)(\deg G - n) + (\gh F + 1)(\gh G +
1)}(G, F),
\nonumber \\
&& (F, GH) = (F, G)H + (-1)^{(\deg F -n)\deg G + (\gh F +1) \gh G}
G(F, H),
\nonumber \\
&& (FG, H) = F(G, H) + (-1)^{\deg G(\deg H -n) + \gh G(\gh H +1) } (F, 
H)G,
\nonumber \\
&& (-1)^{(\deg F - n)(\deg H - n) + (\gh F + 1)(\gh H + 1) } (F, (G,
H))
+ {\rm cyclic \ permutations} = 0,
\label{antibra}
\end{eqnarray}
in $n$ dimensions, where $F, G$ and $H$ are functions on fields and
antifields.

\section*{Appendix B, Dot Product}
\noindent
For superfields $F(\Phi, \Phi^{+})$ and $G(\Phi, \Phi^{+})$,
the following identities are satisfied:
\begin{eqnarray}
&& FG = (-1)^{\gh F \gh G + \deg F \deg G} G F, \nonumber \\
&& d(FG) = dF G + (-1)^{\deg F} F dG, 
\label{FGpro}
\end{eqnarray}
as the usual products.
The graded commutator of two superfields satisfies the following 
identities:
\begin{eqnarray}
&& [F, G] = -(-1)^{\gh F \gh G + \deg F \deg G} [G, F], \nonumber \\
&& [F, [G, H]] = [[F, G], H] 
+ (-1)^{\gh F \gh G + \deg F \deg G} [G, [F, H]].
\label{FGcom}
\end{eqnarray}

We introduce the total degree of a superfield $F$ as 
$|F| = \gh F + \deg F$.
We define the {\it dot product} on superfields as
\begin{eqnarray}
F \cdot G \equiv  (-1)^{\gh F \deg G} FG, 
\label{dotpro}
\end{eqnarray}
and the {\it dot Lie bracket}
\begin{eqnarray}
\lb{F}{G} \equiv (-1)^{\gh F \deg G} [F, G].
\label{dotlie}
\end{eqnarray}
We obtain the following identities
of the {\it dot product} and the {\it dot Lie bracket} from
(\ref{FGpro}), (\ref{FGcom}), (\ref{dotpro}) and (\ref{dotlie}),
\begin{eqnarray}
&& F \cdot G = (-1)^{|F||G|} G \cdot F, \nonumber \\
&& \lb{F}{G} = - (-1)^{|F||G|} \lb{G}{F}, \nonumber \\
&& \lb{F}{\lb{G}{H}} = \lb{\lb{F}{G}}{H}
+ (-1)^{|F||G|} \lb{G}{\lb{F}{H}},
\end{eqnarray}
and
\begin{eqnarray}
d (F \cdot G) \equiv d F \cdot G + (-1)^{|F|} F \cdot d G.
\end{eqnarray}

The {\it dot antibracket} of the superfields $F$ and $G$ is defined as
\begin{eqnarray}
\sbv{F}{G} \equiv (-1)^{(\gh F + 1) (\deg G - n)}
(-1)^{\gh \Phi (\deg \Phi - n) + n}
(F, G),
\label{dotantibra}
\end{eqnarray}
Then the following identities are obtained from the equations
(\ref{antibra}) and (\ref{dotantibra}):
\begin{eqnarray}
&& \sbv{F}{G} = -(-1)^{(|F| + 1 - n)(|G| + 1 - n)} \sbv{G}{F},
\nonumber \\
&& \sbv{F}{G \cdot H} = \sbv{F}{G} \cdot H
+ (-1)^{(|F| + 1 - n)|G|} G \cdot \sbv{F}{H},
\nonumber \\
&& \sbv{F \cdot G}{H} = F \cdot \sbv{G}{H}
+ (-1)^{|G|(|H| + 1 - n)} \sbv{F}{H} \cdot G,
\nonumber \\
&& (-1)^{(|F| + 1 - n)(|H| + 1 - n)} \sbv{F}{\sbv{G}{H}}
+ {\rm cyclic \ permutations} = 0.
\end{eqnarray}
We define the {\it dot differential} as
\begin{eqnarray}
&& \frac{\ld }{\partial \varphi} \cdot F 
\equiv (-1)^{\gh \varphi \deg F}
\frac{\ld F}{\partial \varphi}, \nonumber \\
&&F \cdot \frac{\rd }{\partial \varphi}
\equiv (-1)^{\gh F \deg \varphi}
\frac{F \rd}{\partial \varphi}.
\end{eqnarray}
Then, ~from the equation (\ref{lrdif}), we can obtain the formula
\begin{eqnarray}
\frac{\ld }{\partial \varphi} \cdot F =   
(-1)^{(|F| - |\varphi|) |\varphi|}
F \cdot \frac{\rd }{\partial \varphi}.
\label{lrdiff}
\end{eqnarray}

\vfill\eject
\end{document}